\documentclass[aps,article,floatfix,twocolumn,preprintnumbers,amsmath,amssymb,longbibliography,superscriptaddress,prl]{revtex4-2}

\usepackage{graphicx,xcolor}
\usepackage[caption=false]{subfig}
\begin{document}

\title{Topological Order in an Antiferomagnetic Tetratic}

\author{Daniel Abutbul}
\affiliation{Physics Department, Technion, 32000 Haifa, Israel}

\author{Daniel Podolsky}
\affiliation{Physics Department, Technion, 32000 Haifa, Israel}

\begin{abstract}
We study lattice melting in two dimensional antiferromagnets.  We argue that, for strong enough magnetic interactions, single lattice dislocations are prohibitive due to magnetic frustration.  This leads to a melting scenario in which a tetratic phase, composed of free dislocation {\em pairs} and bound disclinations, separates the solid from the liquid phases.  We demonstrate this phase numerically in a system of hard spheres confined between parallel plates, where spins are represented by the the heights of the spheres. We find that, in the tetratic phase, the spins are as antiferromagnetically ordered as allowed by their spatial configuration. 
\end{abstract}
\maketitle

In a typical antiferromagnet (AF), the N{\' e}el temperature is significantly lower than the melting temperature of the crystal.  For this reason, the magnetic interactions between spins do not play an important role in the crystal melting.  But what would happen if the AF interactions were to be comparable, or even dominant, over the interactions 
responsible for the crystal ordering?  Could this change the nature of the melting transition?

The spin arrangement of an AF depends in detail on the structure of the underlying lattice, making it difficult to envision the lattice melting without destroying the antiferromagnetism at the same time.  Thus, one may naively expect that melting must be a direct transition from an AF-ordered solid to a magnetically-disordered liquid, even when AF interactions are dominant.  

In this Letter, we demonstrate that another possibility exists in two dimensions (2D): the solid and liquid phases could be separated by an intermediate phase, a topologically-ordered tetratic.  This phase relies on strong AF correlations that survive despite the partial melting of the lattice.  We will first give a general argument for this phase, based on a modification of the standard Kosterlitz-Thouless-Halperin-Nelson-Young (KTHNY) picture of defect-mediated melting \cite{Kosterlitz_Thouless_1973,Nelson_Helperin_KTHNY_theory,Young_KTHNY}.  This will be complemented by a numerical demonstration that this phase occurs in an experimentally-realizable microscopic system, a collection of hard spheres confined between parallel plates.

The tetratic phase that we predict has the unusual distinction of having single dislocations bind into free dislocation {\em pairs}, such that double dislocations (of Burgers vector $|{\bf b}|=\sqrt{2}$) are free, while fundamental ($|{\bf b}|=1$) dislocations remain bound.  It is in this sense that the tetratic is topologically-ordered, since higher-charge topological defects proliferate while the fundamental defects do not.

Tetratic phases have been observed in a number of systems \cite{x-atic_phases_of_hard_polygons,Takamichi_multi_colour_MC,kites_tetratic,hard_dimers_tetratic,Hard_square_tetratic?.,random_pinning_tetratic,hard_rectangles_tetratic,rounded_squares_tetratic,binary_solid_solutions_tetratic,melting_scenarios_hertzian_spheres_tetratic,tetratic_FUN_triangleSolid}, including in a closely-related system of Hertzian spheres \cite{Terao_confined_square_hertzian_spheres}.  However, the possibility of topological order, if present, has not been explored.  Evidence for binding of defects was seen in bi-layer systems of interacting particles \cite{Schweigert_double_dislocation_pic}, although the tetratic phase was not observed.  Previous works considered the possibility of topological order \cite{Itamar_molten_af} and AF order \cite{Itamar_molten_af,Carsten_af_liquid} in molten phases, but did not demonstrate these behaviors microscopically.

\begin{figure}[ht!]
    \subfloat{
    \includegraphics[trim=0cm 0cm 0cm 0cm, clip,width=0.45\columnwidth]{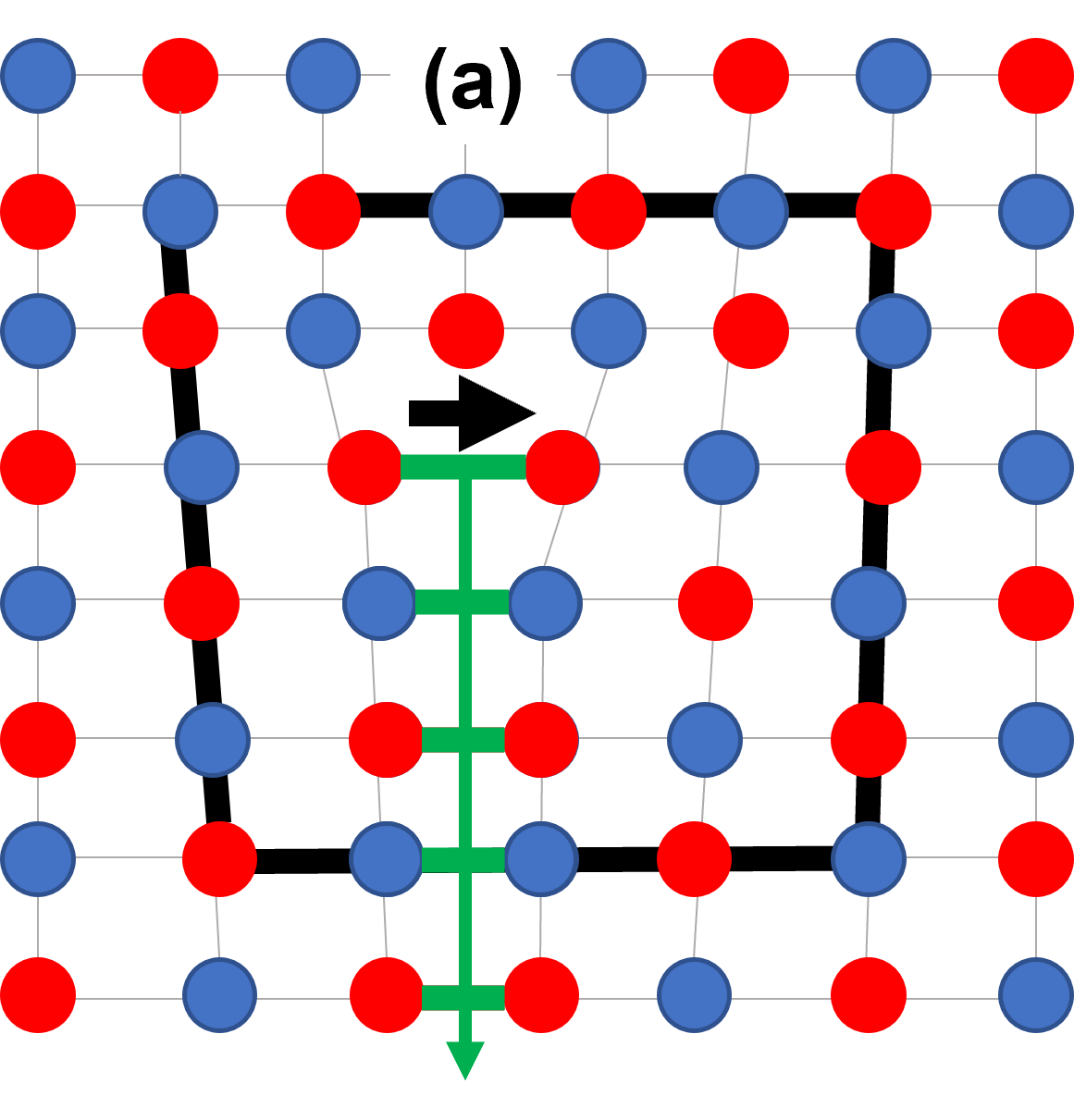}
    }\hfill
    \subfloat{
    \includegraphics[trim=0cm 0cm 0cm 0cm, clip,width=0.45\columnwidth]{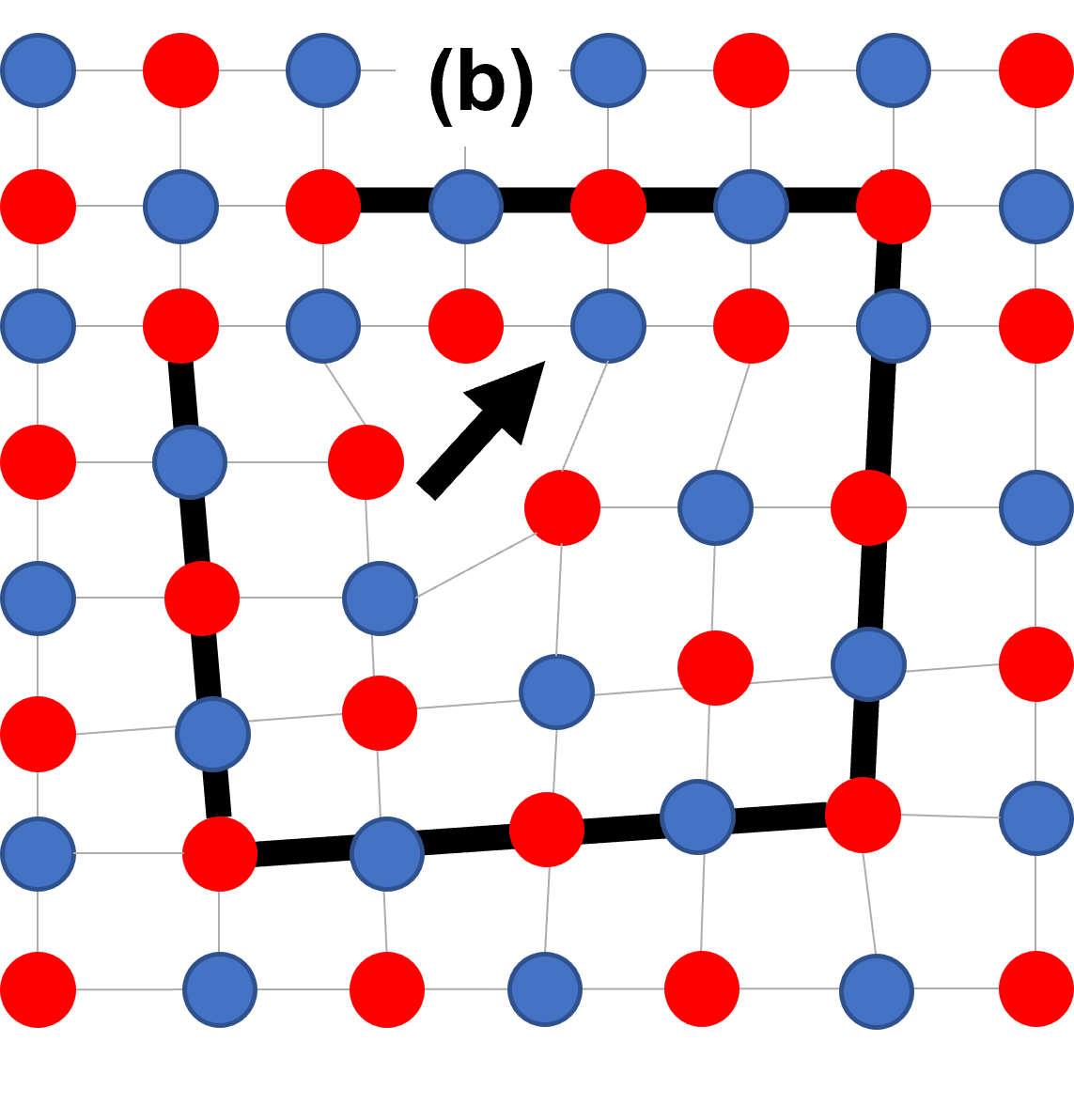}}
 
    \subfloat{
    \includegraphics[trim=0cm 0cm 0cm 0cm, clip,width=0.45\columnwidth]{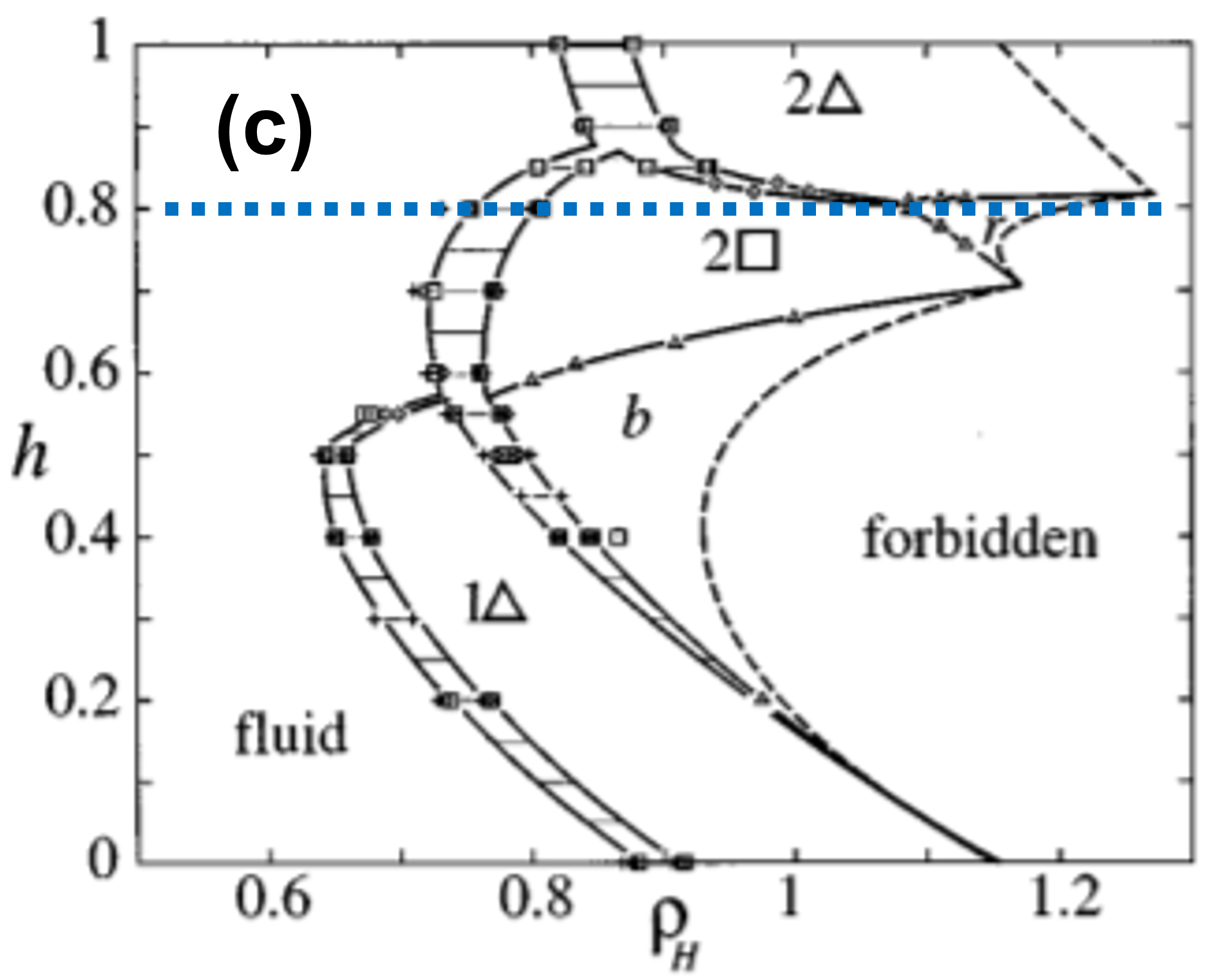}
    }\hfill
    \subfloat{
    \includegraphics[trim=0cm 0cm 0cm 0cm, clip,width=0.45\columnwidth]{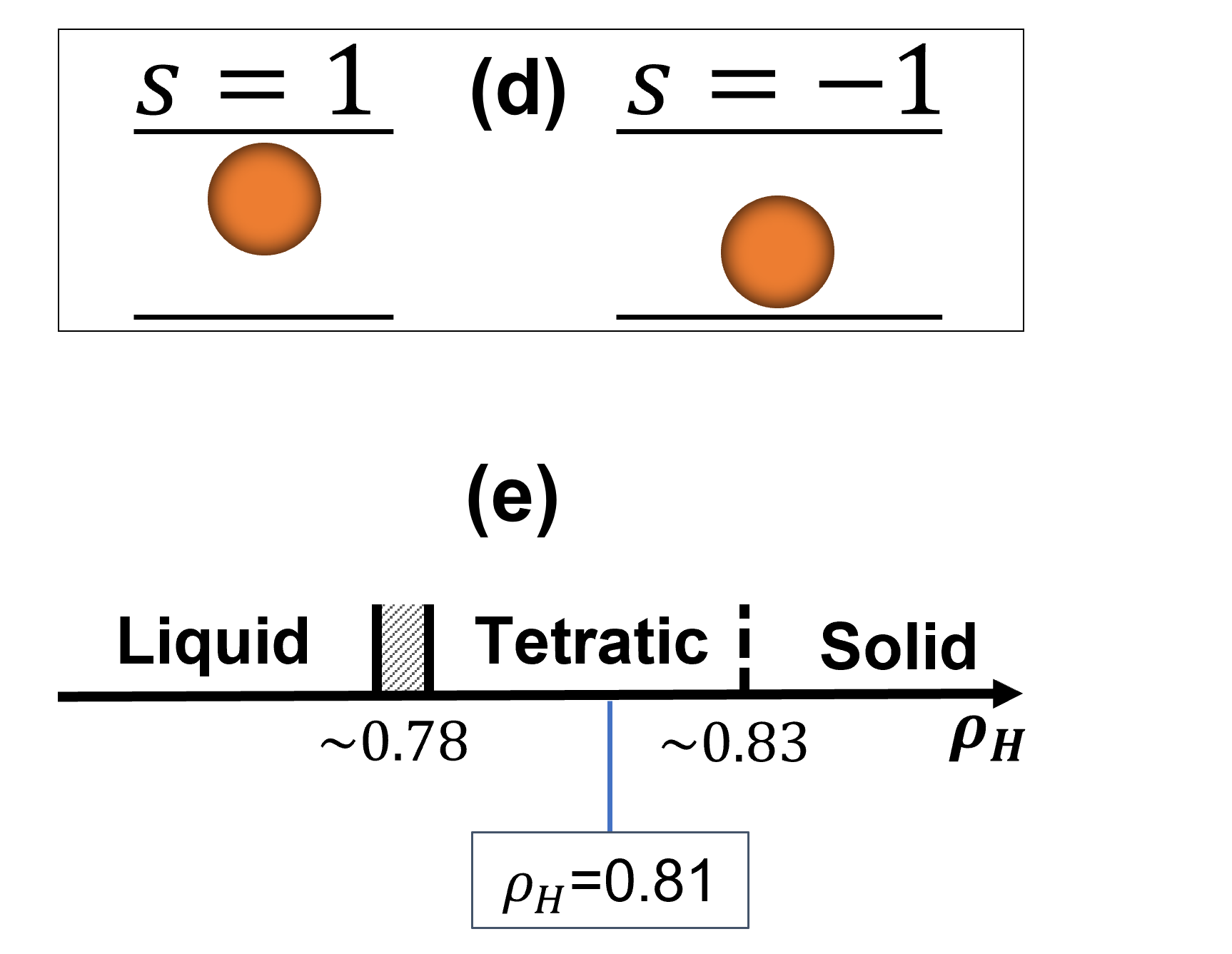}
    }
    \caption{(a) Single dislocation in AF lattice. A string of unsatisfied bonds (green) is emitted from the dislocation core. Burgers circuit is shown by a thick black line. (b) At a double dislocation, no string is created. (c) Phase diagram of confined hard spheres (adapted from \cite{Schmidt_confined_hard_spheres_diagram}). At $h=0.8$ (dotted blue line) the spheres split into upper and lower layers. (d) Spin representation of the spheres: upper (lower) layer is associated with spin $s=1$ ($s=-1$). (e) Phase diagram at $h=0.8$, as obtained in current work. We focus on an intermediate density of $\rho_H=0.81$, which we show is a topologically ordered tetratic with free double dislocations.
    \label{fig:line_of_frustration}}
\end{figure}

We begin by considering the effect of strong AF interactions on the topological defects of a crystal. Consider a system of particles with Ising spins (which take one of two values, $s=\pm 1$), and whose ground state is a N{\' e}el state on a square lattice. As shown in Fig.~\ref{fig:line_of_frustration}(a), a fundamental dislocation of the square lattice breaks the bipartiteness of the lattice and introduces frustration by forcing a semi-infinite string of unsatisfied bonds (across which spins are aligned) to emanate from the dislocation core \cite{Cardy_af_square_double_dislocation,ShokefTopologicalMetamaterialsNature}. This costs an energy that grows linearly with system size, overwhelming the logarithmic free-energy gain associated with the entropy of the dislocation. Hence, the AF background prohibits fundamental dislocations from occurring in isolation.  By contrast, a {\em double} dislocation (Fig.~\ref{fig:line_of_frustration}(b)) has a Burgers vector $|{\bf b}|=\sqrt{2}$ that connects two spins of the same sublattice.  Therefore, it preserves bipartiteness and does not cause frustration \cite{Cardy_af_square_double_dislocation}.  Hence, although double dislocations cost more lattice energy than fundamental dislocations, they do not cost magnetic energy, and they can therefore proliferate when the entropic gain is large enough.  Within the KTHNY scenario, if this were to occur while the disclinations remained bound, a topologically-ordered tetratic would result.

Having described the mechanism underlying the topologically-ordered tetratic, we turn to the question: Does this phase occur in an experimentally-realizable microscopic model?  In order to obtain strong AF interactions, we look for a system in which the AF and crystallization interactions have a common origin.  One such candidate is a collection of hard spheres confined between two parallel plates.    The spheres do not have internal degrees of freedom.  However, when the separation between plates, $H$, is larger than the sphere diameter $\sigma$ but smaller than $2\sigma$, one can think of the height of the spheres relative to the center plane as an effective ``spin''.  In what follows, we will use the terms ``solid'', ``liquid'', and ``tetratic'' to refer to the 2D configuration of the spheres in the plane, without regards to their out-of-plane heights; these will be referred to as their spin.  Although the spheres are non-interacting, other than a hard-core repulsion which forbids overlaps, thermally-induced entropic forces favor configurations in which the spheres pack well. This gives rise to effective AF interactions between the spins, since two nearby spheres can get closer to each other when their heights are different. This system has been studied as a model of AF Ising spins on a hexagonal lattice and can be realized experimentally using colloids \cite{Yair_prl_frustration_ising_colloids,Han_nature_frustration_ising_colloids}.

Note that, since height is a continuous variable, the effective spins are soft. However, at densities in the tetratic and solid phases, we find that the spheres lie very close to the confining plates.  It is then a good approximation to assign the hard spin values $s=\pm 1$, as shown in Fig.~\ref{fig:line_of_frustration}(d).

The phase diagram for this system, as shown in Fig.~\ref{fig:line_of_frustration}(c), was computed in \cite{Schmidt_confined_hard_spheres_diagram} (the possibility of intermediate phases -- tetratic or hexatic -- was not considered).  The phase diagram is characterized by two dimensionless parameters: the normalized density, $\rho_H$, and plate separation, $h$, defined by $\rho_H=\frac{N\sigma^3}{AH}$ and $h=\frac{H}{\sigma}-1$.  Here, $N$ is the number of spheres and $A$ is the total area.  For $h=0$ the system is strictly 2D, such that in the solid phase the spheres arrange in a hexagonal lattice. As the plate separation is increased to $h>0$, the out-of-plane fluctuations increase and so do the effective AF interactions. These can become so strong that they modify the 2D lattice into a bipartite lattice, in order to resolve the magnetic frustration inherent in the hexagonal lattice.  This is an indication that the magnetic and lattice elastic energy scales are comparable in this system.  In our work, we focus on $h=0.8$, where the preferred solid over a wide range of densities is a square lattice (at large values of $\rho_H$, a rhombic structure is preferred due to further-neighbor interactions).

To thermalize the spheres, we use the event-chain Monte Carlo (ECMC) algorithm \cite{Bernard_ECMC}, as it equilibrates quickly and succeeds in escaping local minima \cite{Bernard_two_step_ECMC,two_step_three_sims}. We use our own extension of the 2D $xy$-SEC algorithm, described in Ref.~\cite{Bernard_ECMC}, to three dimensions. In our extension, we apply cyclic boundary conditions in the $\hat{x},\hat{y}$ directions and hard-wall boundary conditions in the $\hat{z}$ direction. When a sphere hits a wall, it bounces from it. See supplementary \cite{Supplementary} for more details.

At $h=0.8$, as the sphere density is reduced starting from $\rho_H=0.9$, we find a sequence of transitions from square lattice solid, to tetratic, to liquid. In order to find the critical densities at the phase transitions, we look for a change of the correlation functions, from algebraic to exponential decay. At the solid to tetratic, the change is in the positional correlations, $g_k(r)$; at the tetratic to liquid, the change is in the bond-orientational correlations, $g_4(r)$ \cite{Terao_confined_square_hertzian_spheres,Helperin_generelized_RG_square_KTHNY,Weikai_two_step_MD}.  These correlation functions are obtained from corresponding complex-valued order parameters, as described next.  

The positional order parameter, evaluated at the location of sphere $\alpha$, is defined with respect to a vector $\bar{\bf k}$ in the reciprocal lattice by 
    $\psi_{\bar{\bf k}}(\alpha)=e^{i\bar{\bf k}\cdot{\bf r}_{\alpha}}$
\cite{Nelson_Helperin_KTHNY_theory}, where ${\bf r}_\alpha=(x_\alpha,y_\alpha)$ is the lateral position of the sphere's center, ignoring its height.   The value of $\bar{\bf k}$ may deviate slightly from the value associated with a perfect lattice of the given density, due to lattice defects \cite{Bernard_two_step_ECMC}.  Instead, we choose $\bar{\bf k}$ at the numerically-evaluated Bragg peak, which maximizes the static structure factor $S({\bf k})=\frac{1}{N} \left|\sum_{\alpha}\psi_{{\bf k}}(\alpha)\right|^2$.
The bond-orientation order parameter of the square lattice is defined by
    $\psi_4(\alpha)=\frac{1}{4}\sum_{\beta\in \mathrm{NN}(\alpha)}{e^{4i\times\theta_{\alpha\beta}}}$ \cite{Schmidt_confined_hard_spheres_diagram},
where $\beta\in \mathrm{NN}(\alpha)$ runs over the 4 nearest neighbors of the sphere $\alpha$, and $\theta_{\alpha\beta}$ is the angle between the bond vector ${\bf r}_\alpha-{\bf r}_\beta$ and the $\hat{x}$ axis. Here, as well, we ignore the sphere heights.
The correlation $g(r)$ of a complex field $\psi({\bf r})$ is defined as the product of the field at points a distance $r$ from each other:
    $g(r)=\int \frac{d^2{\bf r}_\alpha\ d^2{\bf r}_\beta }{A} \frac{\delta\left(|{\bf r}_\alpha-{\bf r}_\beta|-r\right)}{2 \pi r}  \psi^*({\bf r}_\alpha)\psi({\bf r}_\beta)\, .$
As, in practice, we have only samples of the continuous field $\psi({\bf r})$ at discrete positions ${\bf r}_\alpha$, in order to calculate $g(r)$ we use binning of $\psi^*(\alpha)\psi(\beta)$ according to the histogram of separations between pairs of spheres $\left| {\bf r}_{\alpha}-{\bf r}_{\beta} \right|$. The correlation of $\psi=\psi_4$ is denoted by $g_4(r)$ and the correlation of $\psi=\psi_{\bar{\bf k}}$ by $g_k(r)$.

\begin{figure}[ht!]
    \centering
    \includegraphics[trim=0cm 0.5cm 1cm 1.8cm, clip,width=\columnwidth]{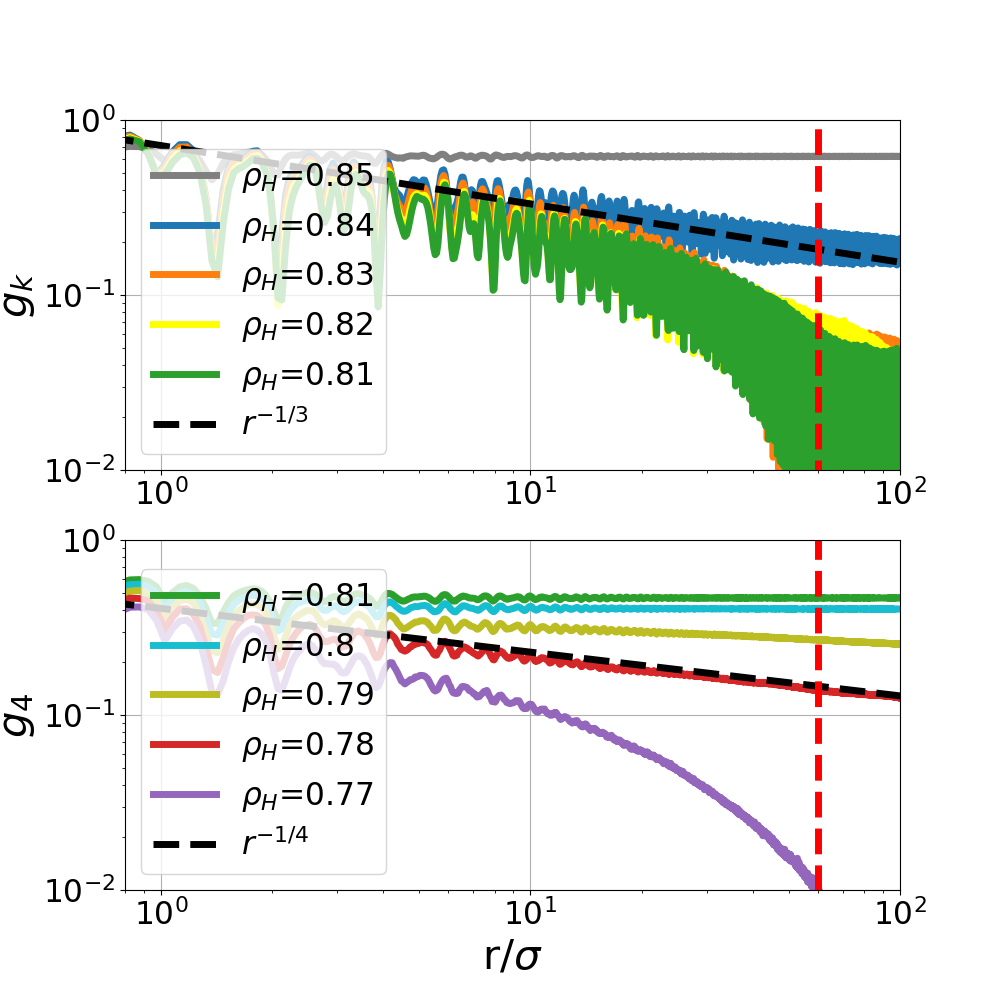}
    \caption{
    Positional ($g_k$) and orientational ($g_4$) correlations for constant plate separation $h=0.8$ and varying density $\rho_H$, for a system with $N=90,000$ spheres (the system size at $\rho_H=0.8$ is $250\sigma \times 250 \sigma \times 1.8\sigma$). Exponential decay of correlation is observed at $\rho_H\leq0.83$ for the positional correlation (upper panel) and $\rho_H\lesssim0.78$ for the bond-orientational correlation (lower panel). The $r^{-1/3}, r^{-1/4}$ bounds from the hexagonal lattice are shown for reference.  The features beyond the dashed red line are associated with finite size effects.
    \label{fig:correlations}}
\end{figure}

 Figure \ref{fig:correlations} shows correlations for various values of $\rho_H$. The change in $g_k(r)$ from power-law to exponential decay happens at $\rho_H\approx0.83$. The onset of exponential decay in $g_4(r)$ happens at $\rho_H\approx0.78$, leaving an intermediate tetratic phase for $0.78\lesssim \rho_H\leq 0.83$. We note here that the tetratic phase should show algebraic decay in $g_4(r)$, whereas we seemingly obtain non-decaying correlations at $\rho_H=0.81$ (as we do in $g_k(r)$ at $\rho_H=0.85$, which lies in the solid phase). Similar results were obtained for the hexatic phase in Refs.~\cite{Weikai_two_step_MD,Bernard_two_step_ECMC}, and may be attributed to very slow algebraic decay.
Note that, for the hexagonal lattice, bounds exist for the decay rate of $g_k(r)$ in the solid ($r^{-1/3}$), and of $g_4(r)$ in the hexatic ($r^{-1/4}$)\cite{Nelson_Helperin_KTHNY_theory,Helperin_generelized_RG_square_KTHNY}. For the square lattice, however, we do not know of such rigorous bounds \cite{Terao_confined_square_hertzian_spheres}. We note that topological order is expected to change these power laws.  In Fig.~\ref{fig:correlations}, we display the standard power laws for the hexatic lattice as a guide to the eye, only.   

We next consider the transition from the tetratic to the liquid, which occurs near $\rho_H\approx 0.78$.  At $\rho_H=0.78$, the orientational correlations look algebraic.  However, at this density, we cannot rule out phase coexistence: we observe bimodal histrograms of the orientational order parameter, $\psi_4$, as well as proliferation of grain boundaries (for more details, we refer to the supplementary \cite{Supplementary}). This suggests that the tetratic to liquid transition is first order \cite{Weikai_two_step_MD}.  A similar scenario, of a continuous solid to {\em hexatic} transition, followed by a first order hexatic to liquid transition, was observed in hard disks in 2D \cite{two_step_three_sims, Bernard_two_step_ECMC,Weikai_two_step_MD}, and in 2D regular polygons \cite{x-atic_phases_of_hard_polygons}.

Having found a tetratic phase, we move on to demonstrate that it is topologically ordered, in the sense of only containing free $|{\bf b}|=\sqrt{2}$ dislocations.
To visualize the dislocation field ${\bf b}({\bf r})$ in the system, we implement a 2D dislocation extraction analysis code (DXA) following Ref.~\cite{Stukowski_dxa}. We first rotate our system by the angle  $-\frac{1}{4} \arg\left(\frac{1}{N}\sum_\alpha{\psi_4(\alpha)}\right)$, thus aligning the average bond orientation with the $\hat{x}$ axis.  We then perform a Delaunay triangulation of the sphere lateral positions.  Then, for each triangle $\alpha\beta\gamma$ of the triangulation, we compute whether it contains a dislocation.  For this, we define the set of separation vectors between nodes of the perfect square lattice, $S'=\left\{(na,ma)|(n,m)\in\mathbb{Z}^2\right\}$, where $a=\frac{2\pi}{|\bar{\bf k}|}$ is the lattice constant. Then, for each edge vector of the triangle, ${\bf \Delta}_{\alpha\beta}={\bf r}_\alpha-{\bf r}_\beta$, we find the nearest vector $\mathbf{\Delta}'_{\alpha\beta}\in S'$ within this set. Finally, the Burgers vector is given by:
    $\mathbf{b}_{\alpha\beta\gamma}=\mathbf{\Delta}'_{\alpha\beta}+\mathbf{\Delta}'_{\beta\gamma}+\mathbf{\Delta}'_{\gamma\alpha}$\,.

\begin{figure}[ht!]
    \subfloat{
        \includegraphics[trim=10.2cm 2cm 10cm 2.5cm, clip,width=0.8\columnwidth]{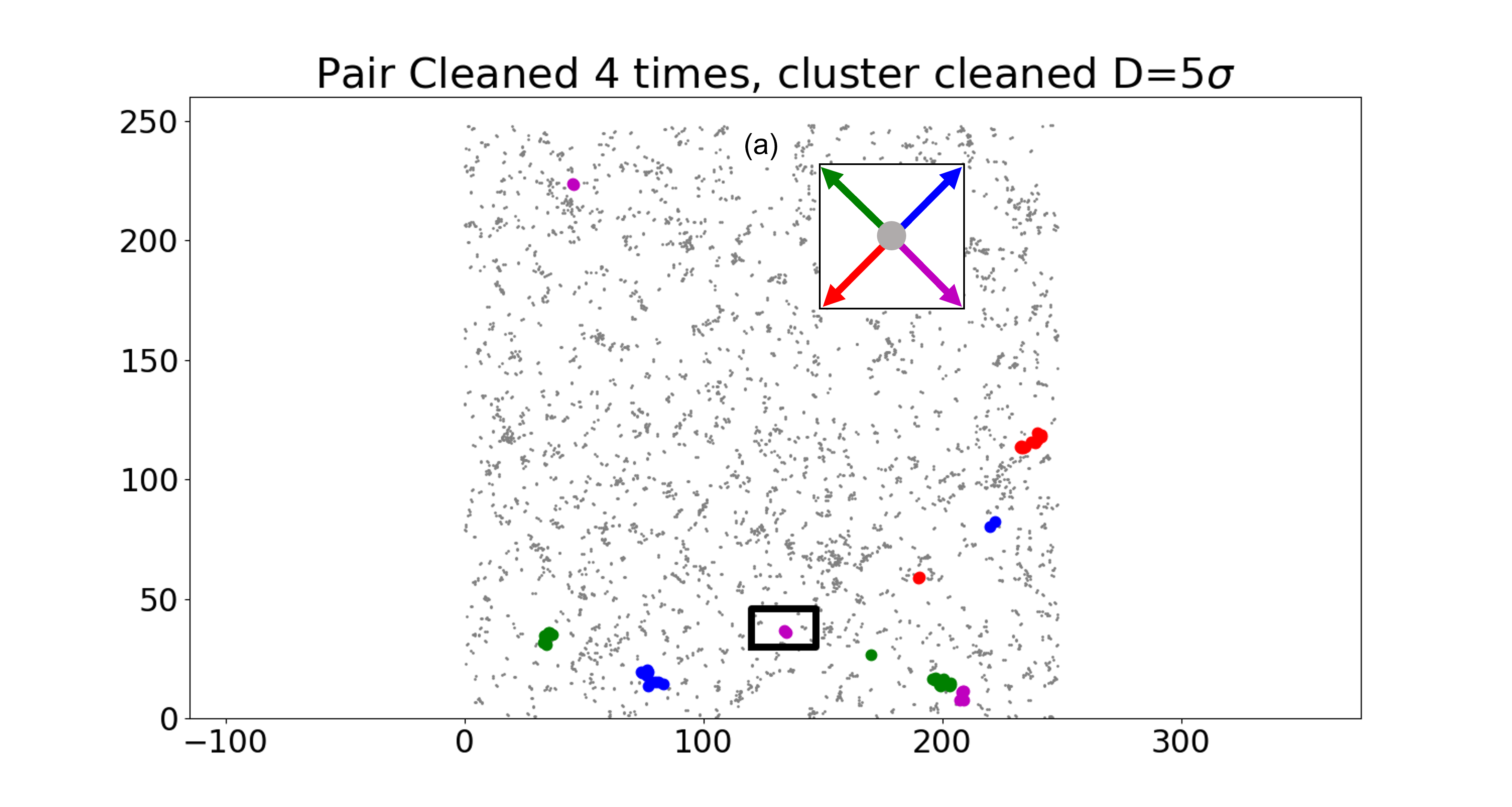}
    }\hfill
    \subfloat{
        \includegraphics[trim=6.5cm 2.5cm 5cm 2.5cm, clip,width=\columnwidth]{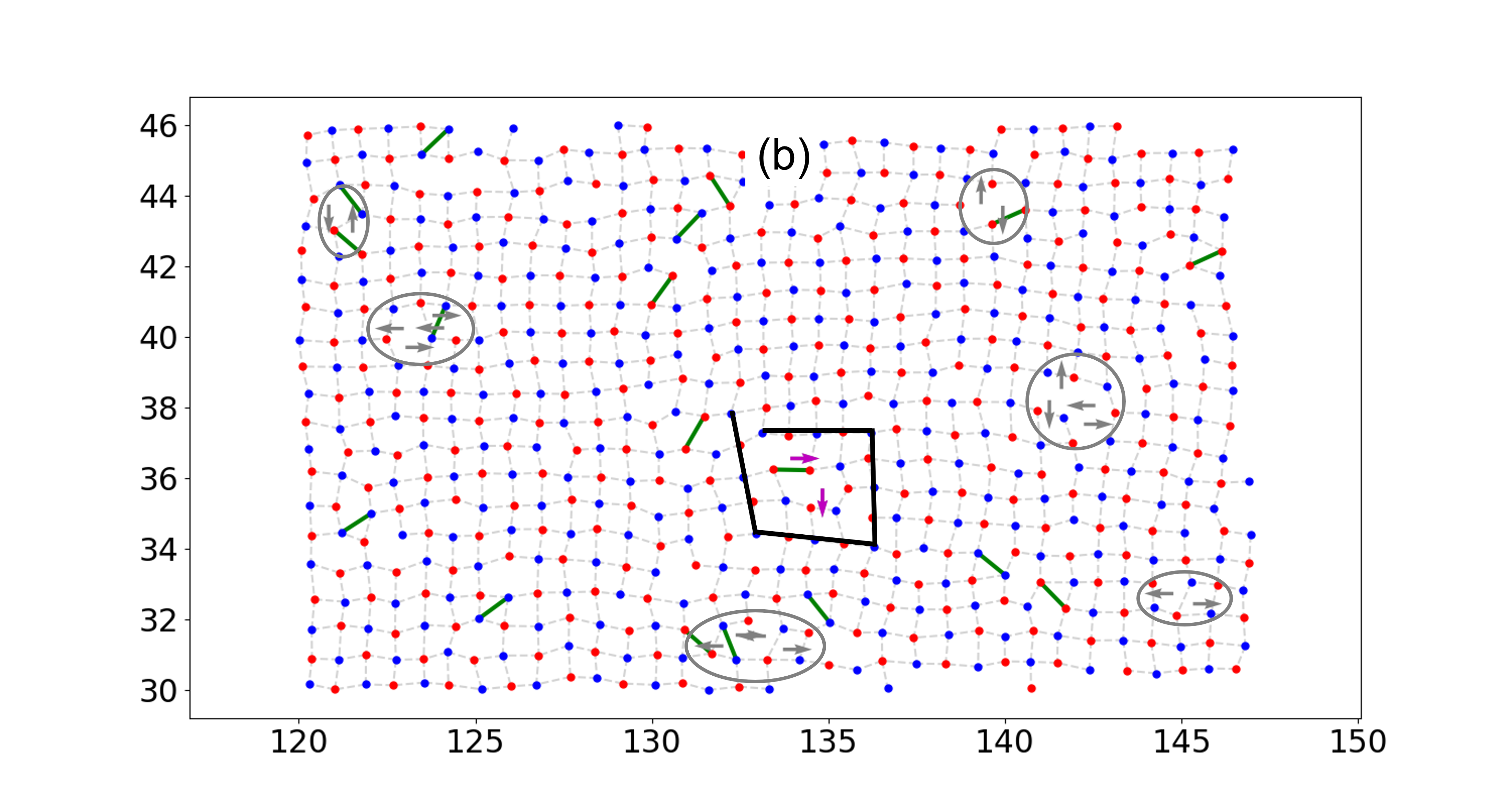}
    }
    \caption{Dislocations for a typical realization of a simulation of $N=90,000$ spheres in the tetratic phase ($h=0.8$ and $\rho_H=0.81$). (a) The Burgers field, calculated using the DXA algorithm \cite{Stukowski_dxa}.  Dislocation-antidislocation pairs and larger neutral clusters are colored in gray.  The remaining clusters,  colored according to the direction of the total Burgers vector of the cluster, $\mathbf{b}_\mathrm{cluster}$, all have magnitude $|\mathbf{b}_\mathrm{cluster}|=\sqrt{2}$. (b)  Zoom-in of the black rectangle in (a).  Here, individual spheres are shown in red (blue) depending on their effective spin $s=+1$ ($s=-1$). A free $\sqrt{2}$ dislocation cluster (magenta arrows) is surrounded by its Burgers circuit; the remaining dislocations (gray arrows), appear in neutral clusters (enclosed by gray ovals).  Dashed gray lines show the bonds of the 4NN graph, used to quantify AF order; green lines are unsatisfied bonds. 
    \label{fig:burger_field}}
\end{figure}

In the tetratic phase, at $\rho_H=0.81$, we indeed find that the free dislocations all have Burgers vector $|{\bf b}|=\sqrt{2}$, as demonstrated in Fig.~\ref{fig:burger_field}(a).  To be more precise, the overwhelming majority of dislocations obtained from DXA have unit Burgers vector.  However, these tend to appear as bound dislocation-antidislocation pairs, or as larger local clusters whose total Burgers vector sums to zero, as illustrated in Fig.~\ref{fig:burger_field}(b). To clean those neutral groups of dislocations, we first remove neighboring dislocation-antidislocation pairs recursively.  We then use a single-linkage agglomerative clustering algorithm \cite{clustering} (see supplementary \cite{Supplementary} for more details).  At the end of this procedure, we find that the resulting clusters are either neutral, or have total $\sqrt{2}$ Burgers vector, indicating that the system is topologically ordered. The free double dislocations are shown in Fig.~\ref{fig:burger_field}(a), with colors according to their orientation. As a check that the cleaning procedure does not bias towards double dislocations, we run the same analysis on mock data consisting of randomly-scattered single dislocations and bound pairs.  We find that, by contrast to the tetratic, in this case nearly all of the final clusters have unit Burgers vector.

\begin{figure}[ht!]
    \centering
    \includegraphics[trim=0cm 0.15cm 0.3cm 0cm, clip,width=\columnwidth]{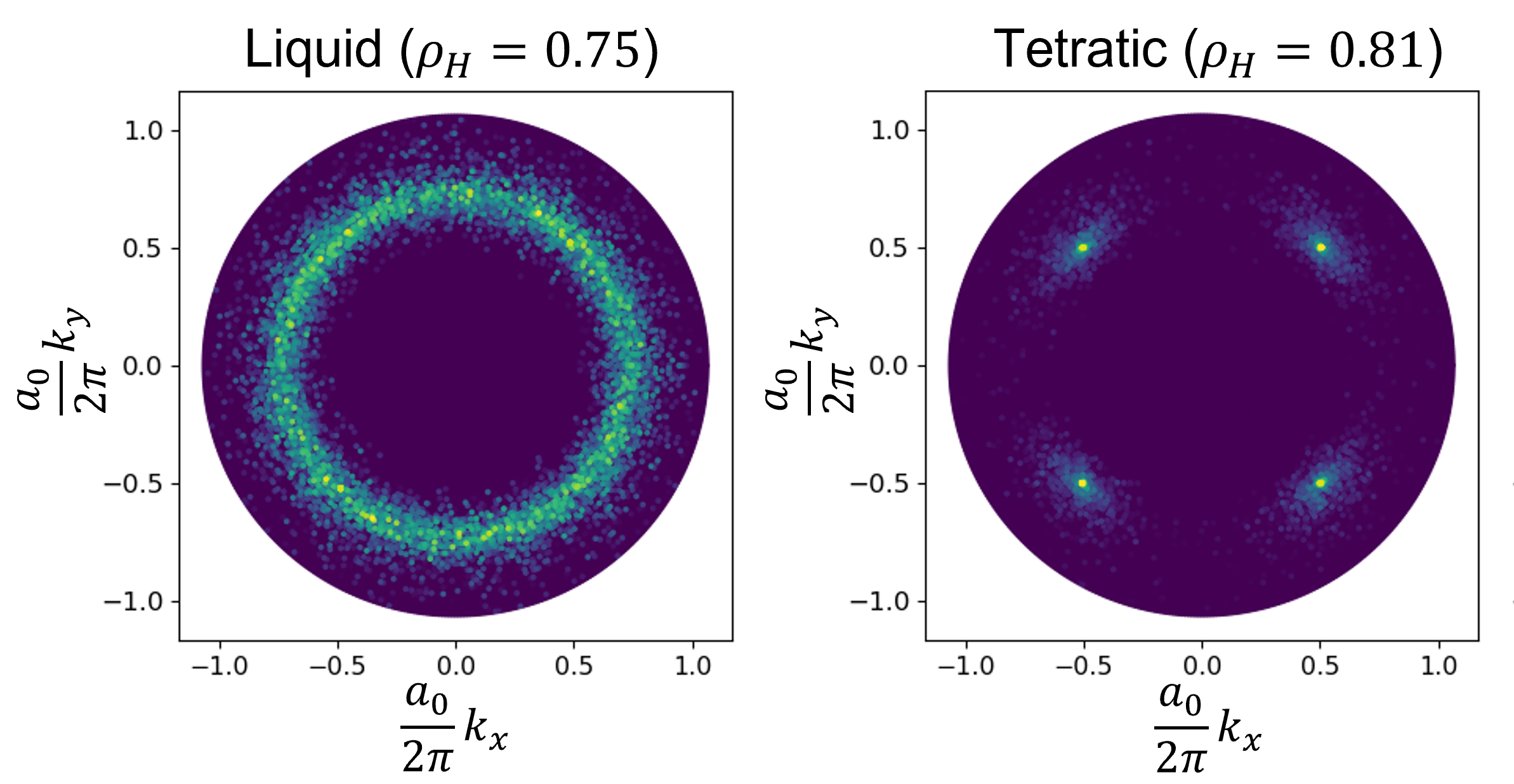}
    \caption{Magnetic structure factor $S^\mathrm{M}({\bf k})$. Fairly sharp peaks, seen in the tetratic, are smeared into a ring in the liquid, indicating strong AF correlations in both phases.\label{fig:MagStruc}}
\end{figure}

Having demonstrated topological order, we next study the antiferromagnetism in the tetratic phase.  As argued above, double dislocations preserve bipartiteness.  Therefore, it may be possible, in principle, for the tetratic phase to have true long-range AF order \cite{Cardy_af_square_double_dislocation,Carsten_af_liquid,Itamar_molten_af}.   It is difficult to give a precise definition of long-range AF order in the tetratic phase.  The usual definition involves the spin-spin correlation function $s_i s_j e^{i{\bf K}\cdot ({\bf r}_i-{\bf r}_j)}$, where ${\bf K}=(\pi/a,\pi/a)$ for a square lattice.  However, in the tetratic phase, this function decays exponentially due to the decay in the positional correlations, regardless of the spin configuration.  Nonetheless, in practice we find fairly sharp peaks in the magnetic structure factor, $S^\mathrm{M}({\bf k})=\frac{1}{N}\sum_{i,j}s_i s_j e^{i{\bf k}\cdot ({\bf r}_i-{\bf r}_j)}$, as shown in Figure ~\ref{fig:MagStruc}.  As the thermodynamic limit is approached, these peaks are expected to eventually spread out into a ring due to the lack of orientational long-range order.   Interestingly, in the liquid phase, we find a ring-shaped maximum in $S^\mathrm{M}$ of radius $\sqrt{2}\pi/a$, indicating that strong AF correlations survive even in the liquid.

We leave aside the question of the possible long-range nature of the AF order, and instead focus on quantifying the degree of antiferromagnetism in the system by comparing it to the ground state of the AF Ising model on a corresponding lattice.  For this, we introduce a criterion for spheres to be considered neighbors, and define the four-nearest-neighbor (4NN) graph, as follows:  Two spheres, $\alpha$ and $\beta$, are connected in the graph iff $\beta$ is one of the four nearest neighbors of $\alpha$ and $\alpha$ is one of the four nearest neighbors of $\beta$.  Figure \ref{fig:burger_field}(b) shows an example of the 4NN graph for a typical configuration in the tetratic phase.  A number of bonds in this graph connect aligned effective spins, {\em i.e.}, they are unsatisfied from the point of view of antiferromagnetism. The unsatisfied bonds in Fig.~\ref{fig:burger_field}(b) are dilute and relatively isolated from one-another.  They occur due to large local fluctuations in sphere positions; near vacancies; or near a bound pair of single-dislocations, which creates a short string of unsatisfied bonds connecting the pair.  All of these effects can result in a local reorganization of the connectivity of the graph, and give rise to a small number of unsatisfied bonds that can be removed by performing local rearrangements of the spheres (or by adding individual spheres in the case of vacancies).   By contrast, if the system had free single dislocations, this would lead to long strings of unsatisfied bonds (or, alternatively, missing bonds) joining faraway dislocations, and to large odd cycles that are not locally removable.  

The degree of antiferromagnetism of a given hard-sphere configuration is captured by the fraction of bonds in the 4NN graph that are unsatisfied, $f_\mathrm{unsat}$. We compare this to the degree of frustration of the 4NN graph, $f_\mathrm{frust}$, which is defined as the minimal fraction of unsatisfied bonds in the graph, obtained when all possible spin configurations are considered. Note that $f_\mathrm{unsat}\geq f_\mathrm{frust}$.  Equivalently, $f_\mathrm{frust}$ is the fraction of unsatisfied bonds in the ground state of the AF Ising model, defined by the Hamiltonian \cite{Robert_ising}:
 $H_{\mathrm{Ising}}=\sum_{\langle \alpha,\beta\rangle}s_\alpha s_\beta$
where the sum runs over nearest neighbors in the 4NN graph.  To find the ground state of $H_{\mathrm{Ising}}$, we consider multiple random initializations of the Ising spins, and slowly anneal from $\beta=0.4$ to $\beta=3$ using the Metropolis algorithm \cite{Robert_ising} (beyond $\beta=3$, the system effectively freezes).

\begin{figure}[ht!]
    \centering
    \includegraphics[trim=2cm 0cm 2.1cm 0cm, clip,width=\columnwidth]{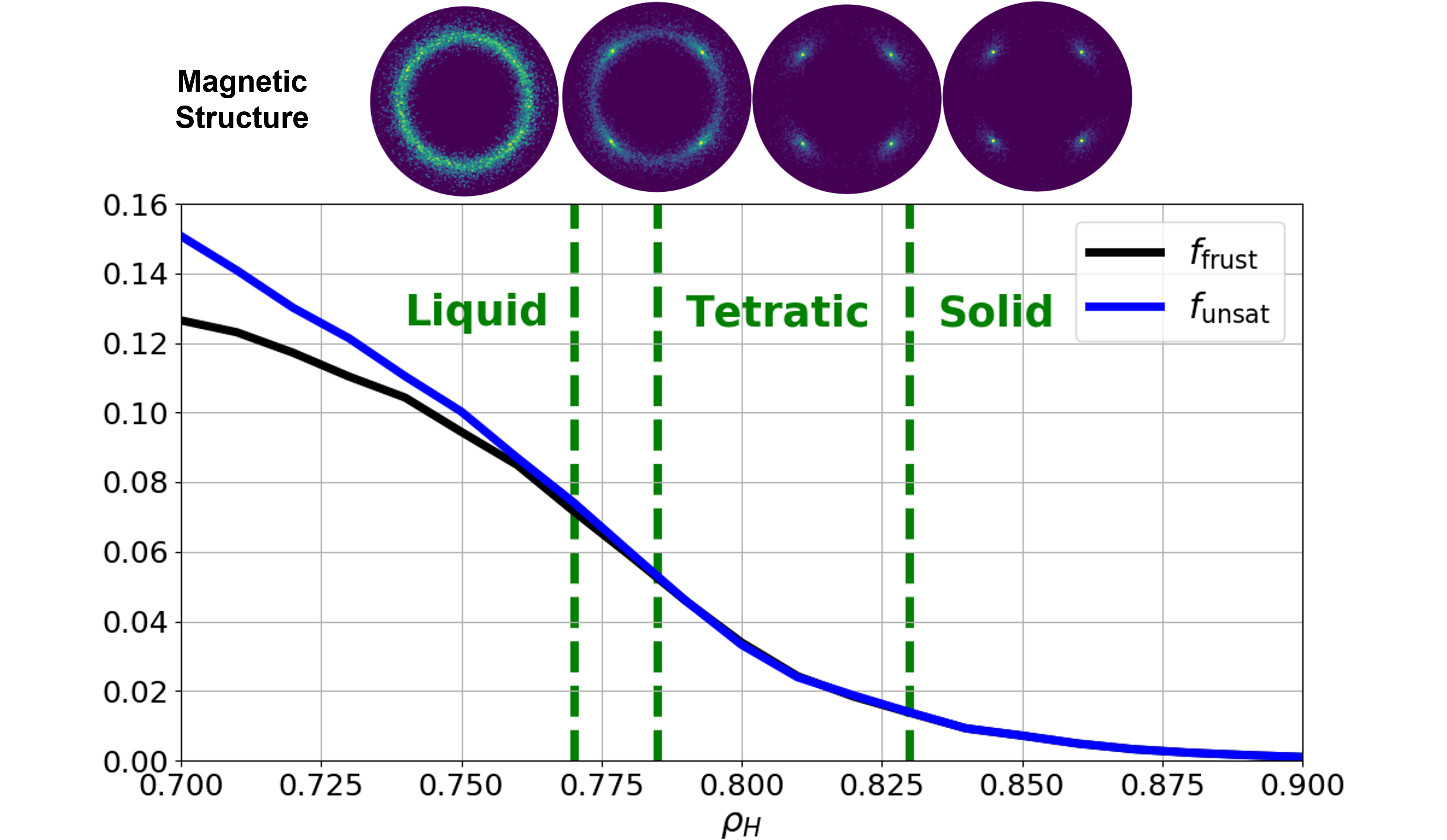}
    \caption{Degree of frustration, $f_\mathrm{frust}$, and fraction of unsatisfied bonds $f_\mathrm{unsat}$, as function of sphere density.  Note that the two closely match throughout the solid and tetratic phases.  Top of the figure: evolution of the magnetic structure factor with $\rho_H$, for $\rho_H=$ 0.75, 0.78, 0.81 and 0.85. \label{fig:Ising}}
\end{figure}

In Figure \ref{fig:Ising}, we see that $f_\mathrm{unsat}$ follows $f_\mathrm{frust}$ in the solid and tetratic phases.
Hence, the effective spins in the hard-sphere system are as AF ordered as allowed by the lattice. This is a very strong result.  In particular, the ability of the effective spins to be in the global minimum of $f_\mathrm{unsat}$, despite the fact that the frustrated lattice is dynamical, is surprising.  It suggests that the lattice dynamics are strongly constrained by the effective spin configuration.   Finally, the large degree of antiferromagnetism confirms our picture, that the dislocations throughout the tetratic phase occur on top of an AF background, with the spins as anti-aligned as possible.  By contrast, if the AF interactions were weak, AF order would be lost well before melting occurs.  

In summary, we have presented a novel phase, a topologically-ordered tetratic, which separates solid and liquid phases in 2D systems with strong AF interactions.  We have shown numerically that this phase is stabilized in a system of hard spheres confined between parallel plates, which may be realized experimentally using colloids.  We find that this phase has fairly sharp magnetic Bragg peaks and saturated AF correlations, despite being partially molten.

{\em Acknowledgments -} We thank Noga Bashan, Yariv Kafri, Dov Levine, Peter Lu, Yair Shokef, and Vincenzo Vitelli for useful discussions, and David Cohen for support in the use of the ATLAS cluster.  We thank the Israel Science Foundation for financial support (grant 1803/18).

\bibliography{topocolloids}

\begin{thebibliography}{32}%
\makeatletter
\providecommand \@ifxundefined [1]{%
 \@ifx{#1\undefined}
}%
\providecommand \@ifnum [1]{%
 \ifnum #1\expandafter \@firstoftwo
 \else \expandafter \@secondoftwo
 \fi
}%
\providecommand \@ifx [1]{%
 \ifx #1\expandafter \@firstoftwo
 \else \expandafter \@secondoftwo
 \fi
}%
\providecommand \natexlab [1]{#1}%
\providecommand \enquote  [1]{``#1''}%
\providecommand \bibnamefont  [1]{#1}%
\providecommand \bibfnamefont [1]{#1}%
\providecommand \citenamefont [1]{#1}%
\providecommand \href@noop [0]{\@secondoftwo}%
\providecommand \href [0]{\begingroup \@sanitize@url \@href}%
\providecommand \@href[1]{\@@startlink{#1}\@@href}%
\providecommand \@@href[1]{\endgroup#1\@@endlink}%
\providecommand \@sanitize@url [0]{\catcode `\\12\catcode `\$12\catcode
  `\&12\catcode `\#12\catcode `\^12\catcode `\_12\catcode `\%12\relax}%
\providecommand \@@startlink[1]{}%
\providecommand \@@endlink[0]{}%
\providecommand \url  [0]{\begingroup\@sanitize@url \@url }%
\providecommand \@url [1]{\endgroup\@href {#1}{\urlprefix }}%
\providecommand \urlprefix  [0]{URL }%
\providecommand \Eprint [0]{\href }%
\providecommand \doibase [0]{https://doi.org/}%
\providecommand \selectlanguage [0]{\@gobble}%
\providecommand \bibinfo  [0]{\@secondoftwo}%
\providecommand \bibfield  [0]{\@secondoftwo}%
\providecommand \translation [1]{[#1]}%
\providecommand \BibitemOpen [0]{}%
\providecommand \bibitemStop [0]{}%
\providecommand \bibitemNoStop [0]{.\EOS\space}%
\providecommand \EOS [0]{\spacefactor3000\relax}%
\providecommand \BibitemShut  [1]{\csname bibitem#1\endcsname}%
\let\auto@bib@innerbib\@empty
\bibitem [{\citenamefont {Kosterlitz}\ and\ \citenamefont
  {Thouless}(1973)}]{Kosterlitz_Thouless_1973}%
  \BibitemOpen
  \bibfield  {author} {\bibinfo {author} {\bibfnamefont {J.~M.}\ \bibnamefont
  {Kosterlitz}}\ and\ \bibinfo {author} {\bibfnamefont {D.~J.}\ \bibnamefont
  {Thouless}},\ }\href {https://doi.org/10.1088/0022-3719/6/7/010} {\bibfield
  {journal} {\bibinfo  {journal} {Journal of Physics C: Solid State Physics}\
  }\textbf {\bibinfo {volume} {6}},\ \bibinfo {pages} {1181} (\bibinfo {year}
  {1973})}\BibitemShut {NoStop}%
\bibitem [{\citenamefont {Nelson}\ and\ \citenamefont
  {Halperin}(1979)}]{Nelson_Helperin_KTHNY_theory}%
  \BibitemOpen
  \bibfield  {author} {\bibinfo {author} {\bibfnamefont {D.~R.}\ \bibnamefont
  {Nelson}}\ and\ \bibinfo {author} {\bibfnamefont {B.~I.}\ \bibnamefont
  {Halperin}},\ }\href {https://doi.org/10.1103/PhysRevB.19.2457} {\bibfield
  {journal} {\bibinfo  {journal} {Phys. Rev. B}\ }\textbf {\bibinfo {volume}
  {19}},\ \bibinfo {pages} {2457} (\bibinfo {year} {1979})}\BibitemShut
  {NoStop}%
\bibitem [{\citenamefont {Young}(1979)}]{Young_KTHNY}%
  \BibitemOpen
  \bibfield  {author} {\bibinfo {author} {\bibfnamefont {A.~P.}\ \bibnamefont
  {Young}},\ }\href {https://doi.org/10.1103/PhysRevB.19.1855} {\bibfield
  {journal} {\bibinfo  {journal} {Phys. Rev. B}\ }\textbf {\bibinfo {volume}
  {19}},\ \bibinfo {pages} {1855} (\bibinfo {year} {1979})}\BibitemShut
  {NoStop}%
\bibitem [{\citenamefont {Anderson}\ \emph {et~al.}(2017)\citenamefont
  {Anderson}, \citenamefont {Antonaglia}, \citenamefont {Millan}, \citenamefont
  {Engel},\ and\ \citenamefont {Glotzer}}]{x-atic_phases_of_hard_polygons}%
  \BibitemOpen
  \bibfield  {author} {\bibinfo {author} {\bibfnamefont {J.~A.}\ \bibnamefont
  {Anderson}}, \bibinfo {author} {\bibfnamefont {J.}~\bibnamefont
  {Antonaglia}}, \bibinfo {author} {\bibfnamefont {J.~A.}\ \bibnamefont
  {Millan}}, \bibinfo {author} {\bibfnamefont {M.}~\bibnamefont {Engel}},\ and\
  \bibinfo {author} {\bibfnamefont {S.~C.}\ \bibnamefont {Glotzer}},\ }\href
  {https://doi.org/10.1103/PhysRevX.7.021001} {\bibfield  {journal} {\bibinfo
  {journal} {Phys. Rev. X}\ }\textbf {\bibinfo {volume} {7}},\ \bibinfo {pages}
  {021001} (\bibinfo {year} {2017})}\BibitemShut {NoStop}%
\bibitem [{\citenamefont {Terao}\ and\ \citenamefont
  {Oguri}(2012)}]{Takamichi_multi_colour_MC}%
  \BibitemOpen
  \bibfield  {author} {\bibinfo {author} {\bibfnamefont {T.}~\bibnamefont
  {Terao}}\ and\ \bibinfo {author} {\bibfnamefont {Y.}~\bibnamefont {Oguri}},\
  }\href {https://doi.org/10.1080/08927022.2012.672740} {\bibfield  {journal}
  {\bibinfo  {journal} {Molecular Simulation}\ }\textbf {\bibinfo {volume}
  {38}},\ \bibinfo {pages} {928} (\bibinfo {year} {2012})},\ \Eprint
  {https://arxiv.org/abs/https://doi.org/10.1080/08927022.2012.672740}
  {https://doi.org/10.1080/08927022.2012.672740} \BibitemShut {NoStop}%
\bibitem [{\citenamefont {Hou}\ \emph {et~al.}(2020)\citenamefont {Hou},
  \citenamefont {Zong}, \citenamefont {Sun}, \citenamefont {Ye}, \citenamefont
  {Mason},\ and\ \citenamefont {Zhao}}]{kites_tetratic}%
  \BibitemOpen
  \bibfield  {author} {\bibinfo {author} {\bibfnamefont {Z.}~\bibnamefont
  {Hou}}, \bibinfo {author} {\bibfnamefont {Y.}~\bibnamefont {Zong}}, \bibinfo
  {author} {\bibfnamefont {Z.}~\bibnamefont {Sun}}, \bibinfo {author}
  {\bibfnamefont {F.}~\bibnamefont {Ye}}, \bibinfo {author} {\bibfnamefont
  {T.~G.}\ \bibnamefont {Mason}},\ and\ \bibinfo {author} {\bibfnamefont
  {K.}~\bibnamefont {Zhao}},\ }\href
  {https://doi.org/10.1038/s41467-020-15723-w} {\bibfield  {journal} {\bibinfo
  {journal} {Nature Communications}\ }\textbf {\bibinfo {volume} {11}},\
  \bibinfo {pages} {2064} (\bibinfo {year} {2020})}\BibitemShut {NoStop}%
\bibitem [{\citenamefont {Wojciechowski}\ \emph {et~al.}(1991)\citenamefont
  {Wojciechowski}, \citenamefont {Frenkel},\ and\ \citenamefont
  {Bra\ifmmode~\acute{n}\else \'{n}\fi{}ka}}]{hard_dimers_tetratic}%
  \BibitemOpen
  \bibfield  {author} {\bibinfo {author} {\bibfnamefont {K.~W.}\ \bibnamefont
  {Wojciechowski}}, \bibinfo {author} {\bibfnamefont {D.}~\bibnamefont
  {Frenkel}},\ and\ \bibinfo {author} {\bibfnamefont {A.~C.}\ \bibnamefont
  {Bra\ifmmode~\acute{n}\else \'{n}\fi{}ka}},\ }\href
  {https://doi.org/10.1103/PhysRevLett.66.3168} {\bibfield  {journal} {\bibinfo
   {journal} {Phys. Rev. Lett.}\ }\textbf {\bibinfo {volume} {66}},\ \bibinfo
  {pages} {3168} (\bibinfo {year} {1991})}\BibitemShut {NoStop}%
\bibitem [{\citenamefont {Wojciechowski}(2004)}]{Hard_square_tetratic?.}%
  \BibitemOpen
  \bibfield  {author} {\bibinfo {author} {\bibfnamefont {K.}~\bibnamefont
  {Wojciechowski}},\ }\href {https://doi.org/10.12921/cmst.2004.10.02.235-255}
  {\bibfield  {journal} {\bibinfo  {journal} {Comp. Meth. Sci. Tech.}\ }\textbf
  {\bibinfo {volume} {10}},\ \bibinfo {pages} {235} (\bibinfo {year}
  {2004})}\BibitemShut {NoStop}%
\bibitem [{\citenamefont {Tsiok}\ \emph {et~al.}(2020)\citenamefont {Tsiok},
  \citenamefont {Fomin}, \citenamefont {Gaiduk},\ and\ \citenamefont
  {Ryzhov}}]{random_pinning_tetratic}%
  \BibitemOpen
  \bibfield  {author} {\bibinfo {author} {\bibfnamefont {E.~N.}\ \bibnamefont
  {Tsiok}}, \bibinfo {author} {\bibfnamefont {Y.}~\bibnamefont {Fomin}},
  \bibinfo {author} {\bibfnamefont {E.~A.}\ \bibnamefont {Gaiduk}},\ and\
  \bibinfo {author} {\bibfnamefont {V.}~\bibnamefont {Ryzhov}},\ }\href@noop {}
  {\bibfield  {journal} {\bibinfo  {journal} {arXiv: Soft Condensed Matter}\ }
  (\bibinfo {year} {2020})}\BibitemShut {NoStop}%
\bibitem [{\citenamefont {Donev}\ \emph {et~al.}(2006)\citenamefont {Donev},
  \citenamefont {Burton}, \citenamefont {Stillinger},\ and\ \citenamefont
  {Torquato}}]{hard_rectangles_tetratic}%
  \BibitemOpen
  \bibfield  {author} {\bibinfo {author} {\bibfnamefont {A.}~\bibnamefont
  {Donev}}, \bibinfo {author} {\bibfnamefont {J.}~\bibnamefont {Burton}},
  \bibinfo {author} {\bibfnamefont {F.~H.}\ \bibnamefont {Stillinger}},\ and\
  \bibinfo {author} {\bibfnamefont {S.}~\bibnamefont {Torquato}},\ }\href
  {https://doi.org/10.1103/PhysRevB.73.054109} {\bibfield  {journal} {\bibinfo
  {journal} {Phys. Rev. B}\ }\textbf {\bibinfo {volume} {73}},\ \bibinfo
  {pages} {054109} (\bibinfo {year} {2006})}\BibitemShut {NoStop}%
\bibitem [{\citenamefont {Avendaño}\ and\ \citenamefont
  {Escobedo}(2012)}]{rounded_squares_tetratic}%
  \BibitemOpen
  \bibfield  {author} {\bibinfo {author} {\bibfnamefont {C.}~\bibnamefont
  {Avendaño}}\ and\ \bibinfo {author} {\bibfnamefont {F.~A.}\ \bibnamefont
  {Escobedo}},\ }\href {https://doi.org/10.1039/C2SM07428A} {\bibfield
  {journal} {\bibinfo  {journal} {Soft Matter}\ }\textbf {\bibinfo {volume}
  {8}},\ \bibinfo {pages} {4675} (\bibinfo {year} {2012})}\BibitemShut
  {NoStop}%
\bibitem [{\citenamefont {Prajwal}\ and\ \citenamefont
  {Escobedo}(2020)}]{binary_solid_solutions_tetratic}%
  \BibitemOpen
  \bibfield  {author} {\bibinfo {author} {\bibfnamefont {B.~P.}\ \bibnamefont
  {Prajwal}}\ and\ \bibinfo {author} {\bibfnamefont {F.~A.}\ \bibnamefont
  {Escobedo}},\ }\href@noop {} {\bibinfo {title} {Novel mesophase behavior in
  two-dimensional binary solid solutions}} (\bibinfo {year} {2020}),\ \Eprint
  {https://arxiv.org/abs/2004.02732} {arXiv:2004.02732 [cond-mat.soft]}
  \BibitemShut {NoStop}%
\bibitem [{\citenamefont {Fomin}\ \emph {et~al.}(2018)\citenamefont {Fomin},
  \citenamefont {Gaiduk}, \citenamefont {Tsiok},\ and\ \citenamefont
  {Ryzhov}}]{melting_scenarios_hertzian_spheres_tetratic}%
  \BibitemOpen
  \bibfield  {author} {\bibinfo {author} {\bibfnamefont {Y.~D.}\ \bibnamefont
  {Fomin}}, \bibinfo {author} {\bibfnamefont {E.~A.}\ \bibnamefont {Gaiduk}},
  \bibinfo {author} {\bibfnamefont {E.~N.}\ \bibnamefont {Tsiok}},\ and\
  \bibinfo {author} {\bibfnamefont {V.~N.}\ \bibnamefont {Ryzhov}},\ }\href
  {https://doi.org/10.1080/00268976.2018.1464676} {\bibfield  {journal}
  {\bibinfo  {journal} {Molecular Physics}\ }\textbf {\bibinfo {volume}
  {116}},\ \bibinfo {pages} {3258} (\bibinfo {year} {2018})},\ \Eprint
  {https://arxiv.org/abs/https://doi.org/10.1080/00268976.2018.1464676}
  {https://doi.org/10.1080/00268976.2018.1464676} \BibitemShut {NoStop}%
\bibitem [{\citenamefont {Prajwal}\ \emph {et~al.}(2021)\citenamefont
  {Prajwal}, \citenamefont {Huang}, \citenamefont {Ramaswamy}, \citenamefont
  {Stroock}, \citenamefont {Hanrath}, \citenamefont {Cohen},\ and\
  \citenamefont {Escobedo}}]{tetratic_FUN_triangleSolid}%
  \BibitemOpen
  \bibfield  {author} {\bibinfo {author} {\bibfnamefont {B.}~\bibnamefont
  {Prajwal}}, \bibinfo {author} {\bibfnamefont {J.-Y.}\ \bibnamefont {Huang}},
  \bibinfo {author} {\bibfnamefont {M.}~\bibnamefont {Ramaswamy}}, \bibinfo
  {author} {\bibfnamefont {A.~D.}\ \bibnamefont {Stroock}}, \bibinfo {author}
  {\bibfnamefont {T.}~\bibnamefont {Hanrath}}, \bibinfo {author} {\bibfnamefont
  {I.}~\bibnamefont {Cohen}},\ and\ \bibinfo {author} {\bibfnamefont {F.~A.}\
  \bibnamefont {Escobedo}},\ }\bibfield  {journal} {\bibinfo  {journal}
  {Journal of Colloid and Interface Science}\ }\href
  {https://doi.org/https://doi.org/10.1016/j.jcis.2021.09.073}
  {https://doi.org/10.1016/j.jcis.2021.09.073} (\bibinfo {year}
  {2021})\BibitemShut {NoStop}%
\bibitem [{\citenamefont
  {Terao}(2013)}]{Terao_confined_square_hertzian_spheres}%
  \BibitemOpen
  \bibfield  {author} {\bibinfo {author} {\bibfnamefont {T.}~\bibnamefont
  {Terao}},\ }\href {https://doi.org/10.1063/1.4822101} {\bibfield  {journal}
  {\bibinfo  {journal} {The Journal of Chemical Physics}\ }\textbf {\bibinfo
  {volume} {139}},\ \bibinfo {pages} {134501} (\bibinfo {year} {2013})},\
  \Eprint {https://arxiv.org/abs/https://doi.org/10.1063/1.4822101}
  {https://doi.org/10.1063/1.4822101} \BibitemShut {NoStop}%
\bibitem [{\citenamefont {Schweigert}\ \emph {et~al.}(1999)\citenamefont
  {Schweigert}, \citenamefont {Schweigert},\ and\ \citenamefont
  {Peeters}}]{Schweigert_double_dislocation_pic}%
  \BibitemOpen
  \bibfield  {author} {\bibinfo {author} {\bibfnamefont {I.~V.}\ \bibnamefont
  {Schweigert}}, \bibinfo {author} {\bibfnamefont {V.~A.}\ \bibnamefont
  {Schweigert}},\ and\ \bibinfo {author} {\bibfnamefont {F.~M.}\ \bibnamefont
  {Peeters}},\ }\href {https://doi.org/10.1103/PhysRevLett.82.5293} {\bibfield
  {journal} {\bibinfo  {journal} {Phys. Rev. Lett.}\ }\textbf {\bibinfo
  {volume} {82}},\ \bibinfo {pages} {5293} (\bibinfo {year}
  {1999})}\BibitemShut {NoStop}%
\bibitem [{\citenamefont {Shamai}\ and\ \citenamefont
  {Podolsky}(2018)}]{Itamar_molten_af}%
  \BibitemOpen
  \bibfield  {author} {\bibinfo {author} {\bibfnamefont {I.}~\bibnamefont
  {Shamai}}\ and\ \bibinfo {author} {\bibfnamefont {D.}~\bibnamefont
  {Podolsky}},\ }\href@noop {} {\bibinfo {title} {Molten antiferromagnets in
  two dimensions}} (\bibinfo {year} {2018}),\ \Eprint
  {https://arxiv.org/abs/1801.08131} {arXiv:1801.08131 [cond-mat.stat-mech]}
  \BibitemShut {NoStop}%
\bibitem [{\citenamefont {Timm}(2002)}]{Carsten_af_liquid}%
  \BibitemOpen
  \bibfield  {author} {\bibinfo {author} {\bibfnamefont {C.}~\bibnamefont
  {Timm}},\ }\href {https://doi.org/10.1103/PhysRevE.66.011703} {\bibfield
  {journal} {\bibinfo  {journal} {Phys. Rev. E}\ }\textbf {\bibinfo {volume}
  {66}},\ \bibinfo {pages} {011703} (\bibinfo {year} {2002})}\BibitemShut
  {NoStop}%
\bibitem [{\citenamefont {Schmidt}\ and\ \citenamefont
  {L\"owen}(1997)}]{Schmidt_confined_hard_spheres_diagram}%
  \BibitemOpen
  \bibfield  {author} {\bibinfo {author} {\bibfnamefont {M.}~\bibnamefont
  {Schmidt}}\ and\ \bibinfo {author} {\bibfnamefont {H.}~\bibnamefont
  {L\"owen}},\ }\href {https://doi.org/10.1103/PhysRevE.55.7228} {\bibfield
  {journal} {\bibinfo  {journal} {Phys. Rev. E}\ }\textbf {\bibinfo {volume}
  {55}},\ \bibinfo {pages} {7228} (\bibinfo {year} {1997})}\BibitemShut
  {NoStop}%
\bibitem [{\citenamefont {Cardy}\ \emph {et~al.}(1983)\citenamefont {Cardy},
  \citenamefont {den Nijs},\ and\ \citenamefont
  {Schick}}]{Cardy_af_square_double_dislocation}%
  \BibitemOpen
  \bibfield  {author} {\bibinfo {author} {\bibfnamefont {J.~L.}\ \bibnamefont
  {Cardy}}, \bibinfo {author} {\bibfnamefont {M.~P.~M.}\ \bibnamefont {den
  Nijs}},\ and\ \bibinfo {author} {\bibfnamefont {M.}~\bibnamefont {Schick}},\
  }\href {https://doi.org/10.1103/PhysRevB.27.4251} {\bibfield  {journal}
  {\bibinfo  {journal} {Phys. Rev. B}\ }\textbf {\bibinfo {volume} {27}},\
  \bibinfo {pages} {4251} (\bibinfo {year} {1983})}\BibitemShut {NoStop}%
\bibitem [{\citenamefont {Meeussen}\ \emph {et~al.}(2020)\citenamefont
  {Meeussen}, \citenamefont {O{\u{g}}uz}, \citenamefont {Shokef},\ and\
  \citenamefont {Hecke}}]{ShokefTopologicalMetamaterialsNature}%
  \BibitemOpen
  \bibfield  {author} {\bibinfo {author} {\bibfnamefont {A.~S.}\ \bibnamefont
  {Meeussen}}, \bibinfo {author} {\bibfnamefont {E.~C.}\ \bibnamefont
  {O{\u{g}}uz}}, \bibinfo {author} {\bibfnamefont {Y.}~\bibnamefont {Shokef}},\
  and\ \bibinfo {author} {\bibfnamefont {M.~v.}\ \bibnamefont {Hecke}},\ }\href
  {https://doi.org/10.1038/s41567-019-0763-6} {\bibfield  {journal} {\bibinfo
  {journal} {Nature Physics}\ }\textbf {\bibinfo {volume} {16}},\ \bibinfo
  {pages} {307} (\bibinfo {year} {2020})}\BibitemShut {NoStop}%
\bibitem [{\citenamefont {Shokef}\ and\ \citenamefont
  {Lubensky}(2009)}]{Yair_prl_frustration_ising_colloids}%
  \BibitemOpen
  \bibfield  {author} {\bibinfo {author} {\bibfnamefont {Y.}~\bibnamefont
  {Shokef}}\ and\ \bibinfo {author} {\bibfnamefont {T.~C.}\ \bibnamefont
  {Lubensky}},\ }\href {https://doi.org/10.1103/PhysRevLett.102.048303}
  {\bibfield  {journal} {\bibinfo  {journal} {Phys. Rev. Lett.}\ }\textbf
  {\bibinfo {volume} {102}},\ \bibinfo {pages} {048303} (\bibinfo {year}
  {2009})}\BibitemShut {NoStop}%
\bibitem [{\citenamefont {Han}\ \emph {et~al.}(2008)\citenamefont {Han},
  \citenamefont {Shokef}, \citenamefont {Alsayed}, \citenamefont {Yunker},
  \citenamefont {Lubensky},\ and\ \citenamefont
  {Yodh}}]{Han_nature_frustration_ising_colloids}%
  \BibitemOpen
  \bibfield  {author} {\bibinfo {author} {\bibfnamefont {Y.}~\bibnamefont
  {Han}}, \bibinfo {author} {\bibfnamefont {Y.}~\bibnamefont {Shokef}},
  \bibinfo {author} {\bibfnamefont {A.~M.}\ \bibnamefont {Alsayed}}, \bibinfo
  {author} {\bibfnamefont {P.}~\bibnamefont {Yunker}}, \bibinfo {author}
  {\bibfnamefont {T.~C.}\ \bibnamefont {Lubensky}},\ and\ \bibinfo {author}
  {\bibfnamefont {A.~G.}\ \bibnamefont {Yodh}},\ }\href
  {https://doi.org/10.1038/nature07595} {\bibfield  {journal} {\bibinfo
  {journal} {Nature}\ }\textbf {\bibinfo {volume} {456}},\ \bibinfo {pages}
  {898} (\bibinfo {year} {2008})}\BibitemShut {NoStop}%
\bibitem [{\citenamefont {Bernard}\ \emph {et~al.}(2009)\citenamefont
  {Bernard}, \citenamefont {Krauth},\ and\ \citenamefont
  {Wilson}}]{Bernard_ECMC}%
  \BibitemOpen
  \bibfield  {author} {\bibinfo {author} {\bibfnamefont {E.~P.}\ \bibnamefont
  {Bernard}}, \bibinfo {author} {\bibfnamefont {W.}~\bibnamefont {Krauth}},\
  and\ \bibinfo {author} {\bibfnamefont {D.~B.}\ \bibnamefont {Wilson}},\
  }\href {https://doi.org/10.1103/PhysRevE.80.056704} {\bibfield  {journal}
  {\bibinfo  {journal} {Phys. Rev. E}\ }\textbf {\bibinfo {volume} {80}},\
  \bibinfo {pages} {056704} (\bibinfo {year} {2009})}\BibitemShut {NoStop}%
\bibitem [{\citenamefont {Bernard}\ and\ \citenamefont
  {Krauth}(2011)}]{Bernard_two_step_ECMC}%
  \BibitemOpen
  \bibfield  {author} {\bibinfo {author} {\bibfnamefont {E.~P.}\ \bibnamefont
  {Bernard}}\ and\ \bibinfo {author} {\bibfnamefont {W.}~\bibnamefont
  {Krauth}},\ }\href {https://doi.org/10.1103/PhysRevLett.107.155704}
  {\bibfield  {journal} {\bibinfo  {journal} {Phys. Rev. Lett.}\ }\textbf
  {\bibinfo {volume} {107}},\ \bibinfo {pages} {155704} (\bibinfo {year}
  {2011})}\BibitemShut {NoStop}%
\bibitem [{\citenamefont {Engel}\ \emph {et~al.}(2013)\citenamefont {Engel},
  \citenamefont {Anderson}, \citenamefont {Glotzer}, \citenamefont {Isobe},
  \citenamefont {Bernard},\ and\ \citenamefont {Krauth}}]{two_step_three_sims}%
  \BibitemOpen
  \bibfield  {author} {\bibinfo {author} {\bibfnamefont {M.}~\bibnamefont
  {Engel}}, \bibinfo {author} {\bibfnamefont {J.~A.}\ \bibnamefont {Anderson}},
  \bibinfo {author} {\bibfnamefont {S.~C.}\ \bibnamefont {Glotzer}}, \bibinfo
  {author} {\bibfnamefont {M.}~\bibnamefont {Isobe}}, \bibinfo {author}
  {\bibfnamefont {E.~P.}\ \bibnamefont {Bernard}},\ and\ \bibinfo {author}
  {\bibfnamefont {W.}~\bibnamefont {Krauth}},\ }\href
  {https://doi.org/10.1103/PhysRevE.87.042134} {\bibfield  {journal} {\bibinfo
  {journal} {Phys. Rev. E}\ }\textbf {\bibinfo {volume} {87}},\ \bibinfo
  {pages} {042134} (\bibinfo {year} {2013})}\BibitemShut {NoStop}%
\bibitem [{\citenamefont {{See Supplemental Material at [URL will be inserted
  by publisher] for more details on the numeric simulations, and more
  observables which help to identify the phases}}()}]{Supplementary}%
  \BibitemOpen
  \bibfield  {author} {\bibinfo {author} {\bibnamefont {{See Supplemental
  Material at [URL will be inserted by publisher] for more details on the
  numeric simulations, and more observables which help to identify the
  phases}}}\ }\href@noop {} {}\BibitemShut {NoStop}%
\bibitem [{\citenamefont {Ostlund}\ and\ \citenamefont
  {Halperin}(1981)}]{Helperin_generelized_RG_square_KTHNY}%
  \BibitemOpen
  \bibfield  {author} {\bibinfo {author} {\bibfnamefont {S.}~\bibnamefont
  {Ostlund}}\ and\ \bibinfo {author} {\bibfnamefont {B.~I.}\ \bibnamefont
  {Halperin}},\ }\href {https://doi.org/10.1103/PhysRevB.23.335} {\bibfield
  {journal} {\bibinfo  {journal} {Phys. Rev. B}\ }\textbf {\bibinfo {volume}
  {23}},\ \bibinfo {pages} {335} (\bibinfo {year} {1981})}\BibitemShut
  {NoStop}%
\bibitem [{\citenamefont {Qi}\ \emph {et~al.}(2014)\citenamefont {Qi},
  \citenamefont {Gantapara},\ and\ \citenamefont
  {Dijkstra}}]{Weikai_two_step_MD}%
  \BibitemOpen
  \bibfield  {author} {\bibinfo {author} {\bibfnamefont {W.}~\bibnamefont
  {Qi}}, \bibinfo {author} {\bibfnamefont {A.~P.}\ \bibnamefont {Gantapara}},\
  and\ \bibinfo {author} {\bibfnamefont {M.}~\bibnamefont {Dijkstra}},\ }\href
  {https://doi.org/10.1039/C4SM00125G} {\bibfield  {journal} {\bibinfo
  {journal} {Soft Matter}\ }\textbf {\bibinfo {volume} {10}},\ \bibinfo {pages}
  {5449} (\bibinfo {year} {2014})}\BibitemShut {NoStop}%
\bibitem [{\citenamefont {Stukowski}(2020)}]{Stukowski_dxa}%
  \BibitemOpen
  \bibfield  {author} {\bibinfo {author} {\bibfnamefont {A.}~\bibnamefont
  {Stukowski}},\ }\bibinfo {title} {Dislocation analysis tool for atomistic
  simulations},\ in\ \href {https://doi.org/10.1007/978-3-319-44677-6_20}
  {\emph {\bibinfo {booktitle} {Handbook of Materials Modeling: Methods: Theory
  and Modeling}}},\ \bibinfo {editor} {edited by\ \bibinfo {editor}
  {\bibfnamefont {W.}~\bibnamefont {Andreoni}}\ and\ \bibinfo {editor}
  {\bibfnamefont {S.}~\bibnamefont {Yip}}}\ (\bibinfo  {publisher} {Springer
  International Publishing},\ \bibinfo {address} {Cham},\ \bibinfo {year}
  {2020})\ pp.\ \bibinfo {pages} {1545--1558}\BibitemShut {NoStop}%
\bibitem [{\citenamefont {Saxena}\ \emph {et~al.}(2017)\citenamefont {Saxena},
  \citenamefont {Prasad}, \citenamefont {Gupta}, \citenamefont {Bharill},
  \citenamefont {Patel}, \citenamefont {Tiwari}, \citenamefont {Er},
  \citenamefont {Ding},\ and\ \citenamefont {Lin}}]{clustering}%
  \BibitemOpen
  \bibfield  {author} {\bibinfo {author} {\bibfnamefont {A.}~\bibnamefont
  {Saxena}}, \bibinfo {author} {\bibfnamefont {M.}~\bibnamefont {Prasad}},
  \bibinfo {author} {\bibfnamefont {A.}~\bibnamefont {Gupta}}, \bibinfo
  {author} {\bibfnamefont {N.}~\bibnamefont {Bharill}}, \bibinfo {author}
  {\bibfnamefont {O.~P.}\ \bibnamefont {Patel}}, \bibinfo {author}
  {\bibfnamefont {A.}~\bibnamefont {Tiwari}}, \bibinfo {author} {\bibfnamefont
  {M.~J.}\ \bibnamefont {Er}}, \bibinfo {author} {\bibfnamefont
  {W.}~\bibnamefont {Ding}},\ and\ \bibinfo {author} {\bibfnamefont {C.-T.}\
  \bibnamefont {Lin}},\ }\href
  {https://doi.org/https://doi.org/10.1016/j.neucom.2017.06.053} {\bibfield
  {journal} {\bibinfo  {journal} {Neurocomputing}\ }\textbf {\bibinfo {volume}
  {267}},\ \bibinfo {pages} {664} (\bibinfo {year} {2017})}\BibitemShut
  {NoStop}%
\bibitem [{\citenamefont {Robert}\ and\ \citenamefont
  {Casella}(2005)}]{Robert_ising}%
  \BibitemOpen
  \bibfield  {author} {\bibinfo {author} {\bibfnamefont {C.~P.}\ \bibnamefont
  {Robert}}\ and\ \bibinfo {author} {\bibfnamefont {G.}~\bibnamefont
  {Casella}},\ }\href@noop {} {\emph {\bibinfo {title} {Monte Carlo Statistical
  Methods (Springer Texts in Statistics)}}}\ (\bibinfo  {publisher}
  {Springer-Verlag},\ \bibinfo {address} {Berlin, Heidelberg},\ \bibinfo {year}
  {2005})\BibitemShut {NoStop}%
\end{thebibliography}%

\end{document}